# Racing to Release: Priority, Congestion, and Community Recognition in Open-Source LLM Ecosystems

**Bin Liu[1,2], Lele Kang[1,2], Jiannan Yang[1,2,*]**

[1]Nanjing University, Laboratory of Data Intelligence and Interdisciplinary Innovation, China
[2]School of Information Management, Nanjing University, China
*Corresponding Author: Jiannan Yang (jnyang@nju.edu.cn)

**ABSTRACT**

Open-source large language models have made platforms such as Hugging Face central hubs for decentralized AI innovation. Yet these ecosystems are shaped not only by collaboration, but also by competition for priority and community attention. Drawing on Hill and Stein's Race-to-the-Bottom framework, this study extends the logic of project potential, maturation, competition, and quality from scientific production to open-source LLM ecosystems, where prominent base models attract concentrated derivative entry under rapid and highly visible platform feedback. Using a large-scale sample of derivative models on Hugging Face, we find that later releases and more crowded competitive environments are both associated with weaker community recognition, even after accounting for differences in model and ecosystem prominence. These findings suggest that competition for priority remains an important organizing force in open-source LLM ecosystems, shaping which derivative innovations receive community recognition.

**KEYWORDS**

Open-Source LLM Ecosystems; Hugging Face; Priority Competition; Race-to-the-Bottom; Community Recognition

## 1 INTRODUCTION

The rapid development of open-source large language models (LLMs) has made platforms such as Hugging Face central hubs for decentralized AI innovation(Jiang et al., 2023). In this ecosystem, developers continuously adapt, fine-tune, and release derivative models based on shared base models. As model release cycles accelerate and community attention becomes increasingly concentrated around emerging model families, developers may face growing pressure to release quickly in order to secure visibility and recognition(Jones et al., 2024). This raises an important question: how does competition for priority shape the downstream recognition of derivative innovation in open-source AI communities?

This question resonates with a long-standing concern in the sociology and economics of science: the tension between allowing a project to mature and establishing priority(Bobtcheff et al., 2017). Building on this tension, Hill and Stein(Hill & Stein, 2025) develop a Race-to-the-Bottom framework in which high-potential projects attract more entry, become more competitive, and are therefore completed more quickly at lower quality. They test this framework in structural biology using data from the Protein Data Bank (PDB), a setting in which projects are relatively discrete and comparable, competition can be identified among contemporaneous teams working on similar structures, project potential can be estimated from ex ante characteristics, and research quality can be measured using objective structure-validation metrics. Their study shows how competition for priority can compress maturation and weaken research quality in a cumulative innovation system.

This study extends that framework to the open-source LLM domain by focusing on a similarly cumulative and competitive innovation setting. On Hugging Face, prominent base models function as shared technological starting points that attract concentrated follow-on entry in the form of derivative releases. As multiple developers build on the same base model under rapid and highly visible platform feedback, derivative innovation becomes shaped not only by model characteristics, but also by short-run competition for attention and recognition. In this respect, open-source LLM ecosystems provide a comparable setting in which priority pressure may influence the timing and downstream reception of innovation. The specific empirical correspondence between the framework and the Hugging Face setting is developed in the following section.

## 2 METHODOLOGY

We adapt the Race-to-the-Bottom framework developed by Hill and Stein (2025), which examines how project potential, competition, and maturation jointly shape research quality in the Protein Data Bank (PDB), to the open-source LLM community on Hugging Face. In their model, high-potential projects attract more entry, become more competitive, and are therefore completed more quickly at lower quality. Translating this logic to the LLM setting, we treat the release of a prominent base model as the opening of a technological opportunity window, and the subsequent derivative models built upon it as competing innovation attempts within the same ecosystem(Vake et al., 2025). Based on the official Hugging Face APIs, we retrieved metadata for 2,556,240 publicly available models on February 3, 2026. Because Hugging Face provides only cumulative popularity indicators such as *likes* and *downloads_all_time*, rather than historical time-series trajectories, we cannot directly observe model quality or popularity at the time of release. To improve comparability, we therefore restrict the analysis to base models released in 2024 and identify the top 100 most downloaded base models in each month. We then trace derivative lineages through repository metadata and retain derivative models released within 365 days after the corresponding base model. After excluding likely automated or spam repositories, defined as models with both likes and downloads below 5, the final sample contains 66,935 valid derivative models nested in 805 active base-model ecosystems.

As shown in Figure 1, following this framework, we define four empirical counterparts to the model's core constructs. For derivative model i, released in week t and built on base model j, **quality** $(Q_{i,j,t})$ is proxied by the log-transformed number of likes received by the derivative model, capturing community-recognized quality on the platform. **Maturation** $(M_{i,j,t})$ is measured as the number of days between the release of the base model and that of the derivative model, reflecting the derivative's release lag within the opportunity window opened by the base model. **Local competition** $(C_{j,t})$ is defined as the log-transformed number of valid derivative models released in the same natural week among models built on the same base model. Finally, following Hill and Stein's emphasis on project potential, we distinguish between **derivative-level potential** $(P_{i,j,t})$ and **base-model-level potential** $(B_j)$, which are measured by the log-transformed number of derivative model's downloads and the base model's downloads separately. In our setting, downloads capture latent attention-generating potential, whereas likes capture realized community recognition. We use the log transformation to reduce the strong right-skewness to make coefficient estimates less sensitive to extreme values.

To improve comparability across lineages, we adopt a fixed 365-day observation window for each base model and retain only derivative models released within one year after the corresponding base model. This window defines which derivative models enter the analytical sample, whereas local competition is measured only at the week in which a focal derivative model is released. We adopt the week-level measure because priority pressure in open-source model release is local and time-sensitive: a daily window is too sparse and noisy, whereas a monthly window is too coarse to capture short-run release congestion. Weekly aggregation therefore provides a practical approximation of the short-run competitive density surrounding a derivative release. We then estimate the following baseline model using derivative models as the unit of analysis:

$$Q_{i,j,t} = \alpha + \beta_1 M_{i,j,t} + \beta_2 C_{j,t} + \beta_3 P_{i,j,t} + \beta_4 B_j + \varepsilon_{i,j,t}, \tag{1}$$

Heteroskedasticity-robust standard errors (HC1) are used throughout. All specifications include base-release-month fixed effects to account for temporal differences across release cohorts. In this way, our design tests whether the competition mechanisms identified by Hill and Stein also characterize derivative innovation within open-source LLM ecosystems.

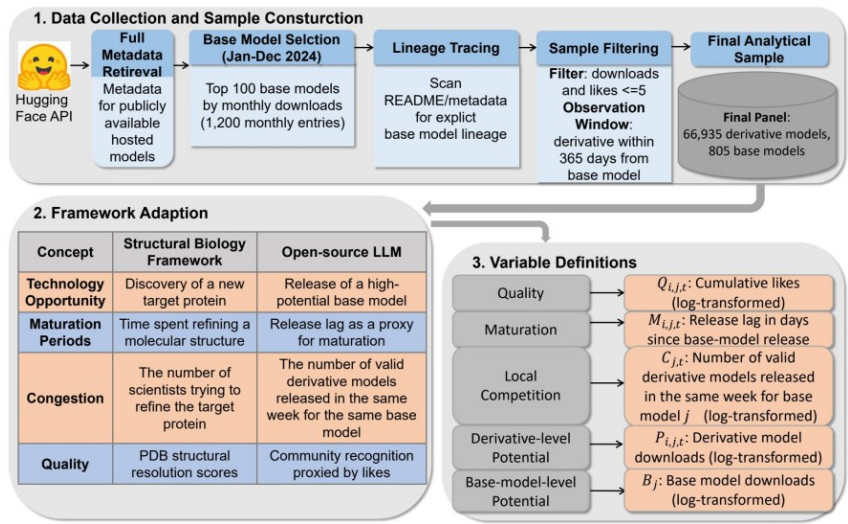

**Figure 1: Methodology for Data Collection and Sample Definition**

## 3 RESULTS

Table 1 reports the baseline regression results for derivative models released within the 365-day observation window. Overall, the baseline specification explains a meaningful share of the variation in downstream community recognition ($R^2 = 0.345$). First, maturation period is negatively associated with derivative-model recognition. The estimated coefficient on maturation period is significantly negative ($p < 0.01$), indicating that, holding model potential constant, derivative models released later within a base-model ecosystem tend to receive fewer community likes. This result suggests that early entry is systematically rewarded in the open-source LLM community, and that delaying release may reduce a model's visibility and recognition within a fast-moving ecosystem. Second, local competition is also negatively associated with derivative-model recognition. The estimated coefficient on local competition is significantly negative ($p < 0.01$), suggesting that derivative models released under more congested local conditions tend to receive lower levels of community recognition. In other words, when multiple derivatives are released in the same ecosystem and within the same short time window, attention appears to become more crowded, weakening the recognition received by any single model. This pattern is consistent with the argument that priority pressure in open-source model release may generate a short-run race-to-the-bottom dynamic. Finally, base-model potential remains positively associated with downstream recognition. Derivative models built upon more highly downloaded base models receive a significant recognition premium, even after controlling for derivative-level characteristics. This suggests that innovation outcomes on Hugging Face are shaped not only by a derivative model's own timing and local competitive environment, but also by the visibility and attractiveness of the ecosystem from which it emerges. Taken together, the estimates reveal a consistent pattern in line with the Race-to-the-Bottom framework: after accounting for both derivative-level and base-model-level potential, shorter release timing and denser local competition are both associated with stronger community recognition penalties

| Term | Coef. (Std. Err.) |
|---|---|
| Maturation Period ($M_{i,j,t}$) | -0.0011*** (0.0001) |
| Local Competition ($C_{j,t}$) | -0.0665*** (0.004) |
| Base-Model-Level Potential ($B_j$) | 0.0597*** (0.004) |
| Derivative-level Potential ($P_{i,j,t}$) | 0.2988*** (0.004) |

| R-Squared ($R^2$) | 0.345 |

**Table 1. Baseline Regression Results for Derivative Models within the 365-Day Observation Window.** *Notes: Robust standard errors (HC1) are reported in parentheses. All specifications include base-release-month fixed effects. *** p<0.01, ** p<0.05, * p<0.1.*

.

To examine whether these relationships are driven by a small number of periods, we further estimate 12 separate monthly cross-sectional regressions. Figure 2 shows that the negative association between maturation period and quality (community recognition) is highly stable over time: the coefficient remains negative in 11 out of 12 months and is statistically significant in most months. By contrast, the coefficient on local competition is more variable, but still negative in 10 out of 12 months, with 7 months reaching statistical significance. The competition effect is especially pronounced during relatively crowded periods, such as August and October.

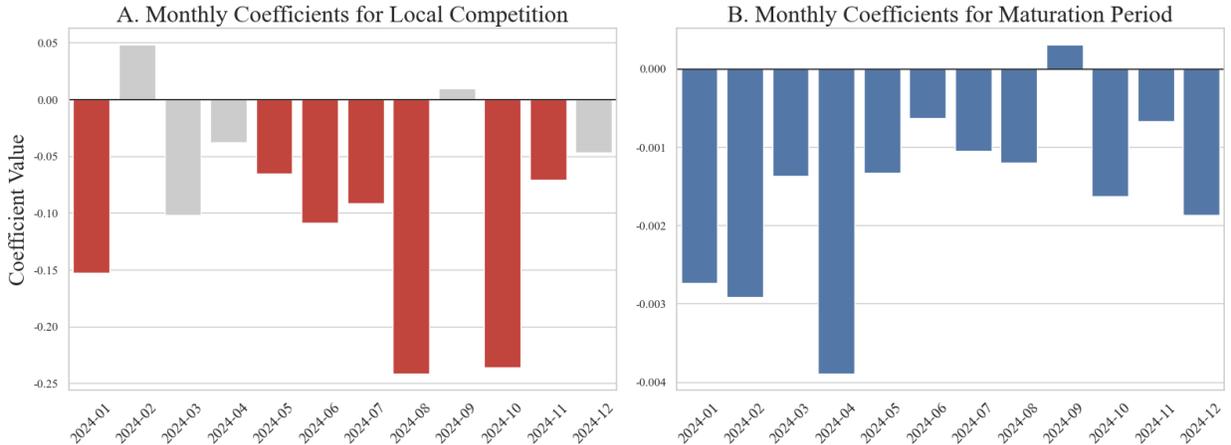

**Figure 2. Monthly coefficient estimates for local competition and maturation period.** A reports the monthly coefficients for local competition, and B reports the monthly coefficients for maturation period. Red and blue bars indicate statistically significant coefficients (p<0.05), while gray ones indicate those not (p>=0.05).

Taken together, these results suggest that the negative relationship between delayed release and community recognition is a persistent feature of derivative-model competition on Hugging Face, while the effect of local competition is directionally consistent but more sensitive to temporal fluctuations in ecosystem congestion. Overall, the evidence supports the view that open-source LLM innovation is shaped by a speed-recognition trade-off, in which both later entry and denser short-run competition are associated with weaker downstream community recognition.

## 4 DISCUSSION AND CONCLUSION

This study extends the Race-to-the-Bottom framework of Hill and Stein (2025) to the open-source LLM community on Hugging Face. Rather than functioning only as collaborative spaces for model reuse, open-source LLM platforms also appear to operate as competitive attention environments(Zhang & Zhang, 2025), in which the timing of release can shape downstream community recognition(Askell et al., 2019; Cave & ÓhÉigeartaigh, 2018). In this sense, the Hugging Face ecosystem exhibits a speed-recognition trade-off similar to the competition dynamics identified in earlier studies of scientific production.

One broader implication of these findings is that openness does not eliminate competitive pressure in AI innovation. On platforms such as Hugging Face, shared base models, rapid derivative entry, and highly visible feedback mechanisms can compress attention cycles and intensify recognition competition. As a result, downstream recognition may reflect not only the substantive merits of a derivative model, but also the timing and density of the release environment in which it appears. This suggests that platform-level recognition structures may play an important role in shaping which innovations gain visibility and traction in open-source LLM ecosystems(Özcan et al., 2026).

At the same time, our interpretation should remain cautious: in this study, quality is approximated through community recognition measured by likes, rather than through direct technical benchmarks(Liang et al., 2023) or safety evaluations. These limitations point to several directions for future research. Benchmark-based measures of technical performance, robustness, and safety would help distinguish more clearly between popularity and substantive model quality. In addition, richer temporal data could improve identification of how recognition evolves immediately after release. Overall, this study shows that competition for priority remains an important organizing force in open-source LLM ecosystems, and that platform recognition may be shaped as much by release timing as by model potential itself.